# Title:

Dual-comb photoacoustic spectroscopy with electro-optic generation for the analysis of gaseous samples

## Author names and affiliations.

Marta Ruiz-Llata, Yuliy M. Sanoyan, Oscar E. Bonilla-Manrique, Pablo Acedo, Pedro Martín-Mateos

Department of Electronic Technology, University Carlos III of Madrid, Spain

## Corresponding author.

Marta Ruiz Llata
Department of Electronic Technology
University Carlos III of Madrid
C/Butarque 15, 28911 Leganés, Spain
Email: mruizl@ing.uc3m.es

## Abstract

In this work we present the design and characterization of a dual comb photoacoustic spectroscopy (DCPAS) set-up for ammonia detection in the near infrared. The system consists of a dual electro-optic (EO) comb generator that generates a multiheterodyne beating signal in the gas sample. The input to the dual EO comb generator is a laser diode tuned to a fixed wavelength within an absorption feature of ammonia (around 1531.6 nm ≡ 6529 $cm^{-1}$) and we show how the dual comb allows to perform PAS measurements and resolve the absorption features with high spectral resolution. We present results of the ammonia absorption line profile reconstruction with a bandwidth of 1 $cm^{-1}$ and a resolution of 0.08 $cm^{-1}$. Moreover, we show that dual comb technique based on electro-optic generation maximally simplifies the optimization of the multiheterodyne signal according the characteristic of the photoacoustic detection module. We present results using a resonant gas cell (pipe shape) and we show how easily the dual comb optical source is adjusted to generate multiherodyne beating tones within the band of resonance of the gas cell.

## 1. Introduction

Photoacoustic spectroscopy (PAS) has been used in different applications for trace gas-sensing due to several features such as high selectivity, low cost, linear response, and wide dynamic range compared to other absorption spectroscopic techniques [1]. One of its key advantages is the measuring system compactness that allows to attain high sensitivity levels using very small sample volumes [2-4]. Moreover, the gas absorption photoacoustic detection module, typically a standard microphone [1], a quartz tuning fork in quartz enhance PAS (QEPAS) [3-6], or a cantilever with interferometric readout [7-9], is excitation wavelength independent, making the PAS technique particularly suitable for broadband detection and a very interesting solution for the mid-infrared region of the spectrum, where fundamental molecular vibration and rotation absorption lines are found. Broadband spectral measurement of gas absorption enables accurate measurements in complex systems where interfering substances may be present or environmental parameters cannot be controlled.

The typical methods used in optical transmission based measurements for broadband gas spectroscopy [10] can be also applied for PAS. One of the preferred approaches is the combination of wavelength tuning and wavelength modulation of a semiconductor laser diode [11], but this approach is limited by the tuning range of the semiconductor laser, and the wavelength modulation depth needs to be optimized to the shape of the absorption line, which depends on the temperature and pressure of the gas sample. Another approach is the use of multiple laser sources [12], however those need to be selected to the absorption features of the selected target species and coupling the different optical beams into the detection module may result in a complicated optical set-up.

On the other hand, Fourier-transform infrared (FTIR) spectrometers, in combination with broadband optical sources, have been demonstrated as a powerful tool for PAS [13-15], as it permits wide spectral coverage and very high spectral resolution if, for example, an optical frequency comb (OFC) is used as the light source [16]. However, this technique requires mechanically stable set-ups for optical scanning, which compromise the spectral bandwidth and the acquisition speed. As an alternative to FTIR, dual-comb spectroscopy (DCS) has emerged as revolutionary tool for broadband spectroscopy with unprecedented accuracy and precision, without requiring bulky, alignment sensitive, mechanical parts [17].

DCS, also known as multiheterodyne spectroscopy, relies on two OFCs. An OFC is an optical source with a broad spectrum consisting on evenly spaced, tightly phase-locked, optical frequencies. In DCS systems, an OFC passes thought the gas sample and the transmitted light is heterodyned with a second OFC (the local oscillator OFC) at the detection photodiode. As the local oscillator OFC is coherent with the measurement OFC, but with a slightly different optical frequency spacing, the photodiode signal originates a new comb in the RF domain that contains the spectral information of the measurement OFC. Alternatively, both OFC can be made to propagate simultaneously through the gas sample. In this case, the interference of the two OFC by the absorbing medium can generate a photoacoustic signal if the line-by-line beating of the OFCs occurs at acoustic frequencies. This broadband absorption detection scheme has been recently demonstrated for solid and gas samples in a technique known as dual comb photoacoustic spectroscopy (DCPAS) [18][19].

State of the art dual comb generation technologies are mode-locked fiber lasers [21], microresonator-based frequency combs [22], optical frequency combs based on Quantum Cascade Lasers [23] and electro-optic modulation of CW semiconductor lasers (EO combs) [24-

28]. This later approach has shown a growing interest in the recent years for its advantageous features of intrinsically high mutual coherence, efficient use of the optical power with relative narrow spectral coverage and high power per comb line, unbeatable dual comb parameters configuration capability, and technological maturity in the near-infrared [29]. Recent examples of prospect spectroscopy based applications using dual EO combs include space mounted greenhouse gas detection LIDARs [30], and hyperspectral imaging systems [31] and sensors [32]. In this work we investigated the performance of an electro-optic dual comb generator-based DCPAS system for the detection of ammonia absorption features in the near-infrared using a resonant acoustic detection module. We show that the unique properties of the EO comb generator and their flexibility in configuring the mapping of the optical spectrum to the acoustic spectrum allow resolving spectral signatures with detailed spectral information.

## 2. Material and methods

### 2.1. Measurement set-up

Figure 1 depicts the block diagram of our experimental set-up. It is composed by a dual OFC generated by electro-optic modulation of a CW semiconductor laser diode (LD) that is used as excitation to a standard photoacoustic detection module.

The purpose of the dual EO comb generator is to provide broadband detection of target gases using any standard or custom photoacoustic detection module. According to the background theory of DCS and DCPAS, that can be found on references [17] and [18, 19] respectively and are summarized in the next section, the target gas sample will be excited by a signal whose optical spectrum can be defined by a set of frequencies $\nu_n = \nu_0 + n \cdot \Delta\nu$, with n = 0, ±1, ±2…, and the photoacoustic response will have spectral components at frequencies $f_n = f_0 + n \cdot \Delta f$, being the amplitude of each n tone proportional to the absorption of the gas sample at the frequency $\nu_n$.

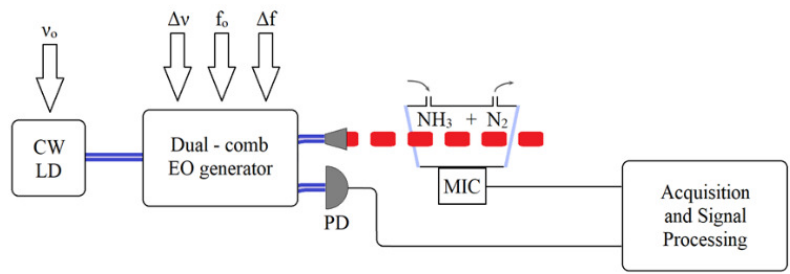

Figure 1. Block diagram of the experimental set-up. $\nu_0$ and $\Delta\nu$ are the free parameters of the optical excitation spectrum. $f_0$ and $\Delta f$ are the free parameters of the generated acoustic signal.

In our set-up the photoacoustic detection module is a gas cell consisting of an aluminum cylinder resonator (a pipe), with a 5-mm input diameter and a length of 88 mm, terminated with two cylindrical buffer volumes as acoustic filters. The acoustic detector is a microphone (Knowles FG-23329), placed at the center of the resonator, and connected to an amplifier (Femto DHPVA-101). In this work, the acquisition and signal processing is based on a lock-in amplifier (Zurich Instruments HF2LI) with two inputs, one for the acoustic signal and the other for a reference of the optical beat note signal from the dual EO comb generator using a photodiode (ThorLabs PDA10CF-EC). The measurements were made with a gas flow of 60 ml/m of ammonia at 5% concentration though the gas cell at atmospheric pressure.

## 2.2. Dual-comb photoacoustic spectroscopy (DCPAS) method

In DCPAS, as represented in figure 2, a multiheterodyne signal is generated using two mutually coherent OFCs. An OFC is a pulsed light source with an optical spectrum consisting of many equidistant monochromatic tones (comb teeth). The optical characteristics of an OFC are given by the pulse repetition rate, which also sets the comb teeth separation frequency ($f_{REP}$), and the carrier envelope offset frequency ($f_{CEO}$), so the optical frequency of any comb line N, being N natural number, within the optical span of the OFC, can be expressed by $f_N = f_{CEO} + N \cdot f_{REP}$. The mutiheterodyne signal is generated when two OFCs with slightly different $f_{REP}$ are combined and propagated through the sample. An additional condition is that the two OFCs must be mutually coherent, that is to say, there is a constant mismatch between the $f_{CEO}$ of the two OFCs. As a result, the optical intensity of the combined combs contains a multiheterodyne term corresponding to the beat notes of each (increasingly offset) pair of comb teeth. In the optical domain, because the difference of the repetition rate between the OFCs ($\Delta f_{REP}$) is very small compared to the repetition rate ($f_{REP}$), we can assume the same optical frequency $v_n$ for each pair of comb teeth. In the temporal domain each optical tone with frequency $v_n$ is amplitude modulated at the frequency $f_n$, being this the beating frequency of the corresponding dual comb teeth.

For DCPAS, the optical span of the dual comb is centered around the absorption features of the target molecules, while the multiherodyne spectrum spam is centered within the acoustic bandwidth of the detection module (i.e. within the bandwidth of the microphone in the case of a non-resonant gas cell). Then, when the gas sample absorbs light from the dual comb source it produces a photoacoustic response that reproduces the absorption profile of the sample, mapping the absorption at each optical frequency $v_n$ to a different acoustic frequency $f_n$.

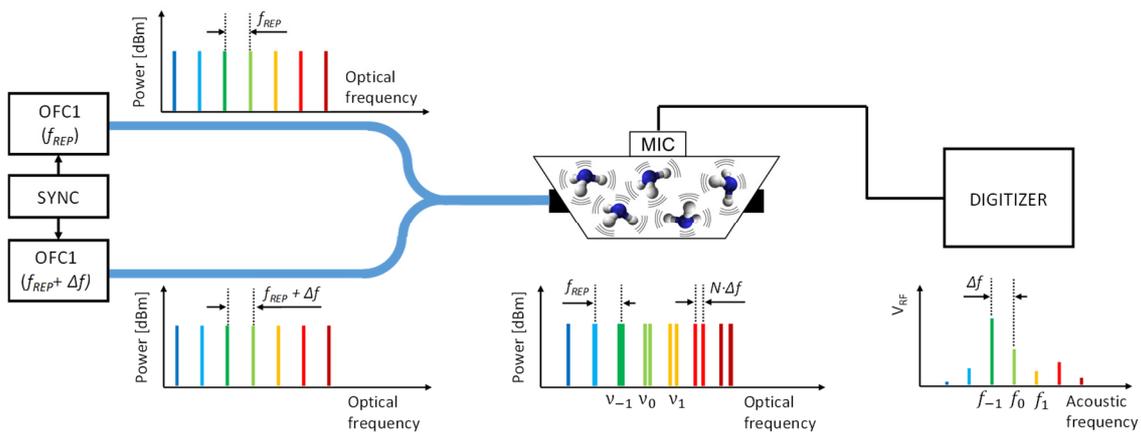

Figure 2. Principle of DCPAS: Two OFCs that are very similar in their optical frequencies ($\Delta f \ll f_{REP}$) are heterodyned within the gas sample. As OFC1 and OFC2 are synchronized (their $f_{CEO}$ are locked), then acoustic tones are generated with a line spacing equal to $\Delta f$ and with an amplitude that reproduces the absorption profile of the sample.

In this paper we focus on the investigation of the EO comb generation technique for broadband PAS due to its unique ability to easily adjust the operation parameters depending on the specific application and on the acoustic detection module. Electro-optic comb generators have proved important advantages over conventional mode-locked laser-based architectures. Most relevant advantages for DCPAS are exquisite simplicity, tailored spectral coverage and optical resolution ($v_0$, $\Delta v$), frequency agility, high acquisition rates, and the inherently high mutual coherence between combs, that enables mHz intermode beat note linewidths [19,29] and virtually unmatched spectral compression ratios ($\Delta v / \Delta f$). The details of the dual EO comb generator used in our experiments are described in section 3. It will allow the configuration of: $v_0$, the central optical frequency of the excitation light source; $f_0$, the frequency of the acoustic signal generated by the absorption of the gas sample at $v_0$; $\Delta v$, the optical frequency difference between consecutive tones of the excitation light source (optical resolution); and $\Delta f$, the frequency difference between acoustic tones.

## 3.- System implementation and characterization

### 3.1.- Electro-optic dual comb generator

The dual comb is obtained from a single near-infrared continuous wave DFB laser diode (LD Laser in Fig. 3) whose output is split to generate two OFCs and then recombined. This guarantees intrinsically high mutual coherence since both OFCs are generated from the same CW LD. The central optical frequency of the resulting dual comb ($v_0$) is fixed by the LD wavelength, which can be selected by tuning the LD current and temperature.

Each OFC is generated using one electro-optic phase modulator (PM). Each PM is driven by a different signal generator (SG12 and SG22 in Fig. 3), both set to a very close frequency in the GHz range. The phase modulators generate lower and upper sidebands (comb teeth) around the input optical tone, depending the number of modes and their relative amplitudes on the modulation index (RF power). This configuration is the simplest and most straight forward method for EO comb generation and provides OFCs with a reduced number of teeth and poor flatness, however better comb characteristics in terms of number of comb teeth and flatness can be achieved cascading EO modulators [29]. In our set-up the number of comb teeth we generate is in the order of 10, this number has been demonstrated as an optimum compromise between the quality of the concentration retrieval targeting a single spectral signature and the energy per tone, taking into account a limit on the available total power and the required signal to noise ratio [30].

Electro-optic OFCs have other important advantages for DCPAS. One of the advantages is the relatively large mode spacing (compared with mode-locked laser OFCs) and the flexibility to set this parameter by adjusting the modulation frequency of the PMs. In our set-up we make the frequency of SG21 equal to $\Delta v$, which is the desired optical spectral resolution, and the frequency of SG22 equal to $\Delta v + \Delta f$, so that each comb tooth pair provides a single independent beating note with a frequency proportional to $\Delta f$.

The flexibility of OE comb generation is extended not only to the optical to acoustic conversion bandwidth, but also to the optical to acoustic central frequency conversion, that can be adjusted independently. This is done by optical frequency shifting the OFCs using an acousto-optic modulator (AOM) placed at each branch of the dual-comb generator, each driven by a different signal generator (SG11 and SG21 in Fig. 3) set to a very close frequency in the tens of MHz range. In our set-up we make the frequency of SG11 equal to 40 MHz, which is the working frequency of the AOMs we used, and the frequency of SG21 equal to 40 MHz + $f_0$. Because the frequency shift is very small compared with the laser frequency (40 MHz compared with $v_0$ = 195.7 THz), we can assume the same optical central frequency at the output of the dual comb generator ($v_0$) and a beating term that generates a photoacoustic response at the frequency $f_0$, that is set according the detection module we use.

Mutual coherence between combs is obtained because they share the same optical input and the same RF time base, since all signal generators share the same clock signal. Mutual coherence is a requirement to spectrally resolve the photoacoustic comb, so that each dual comb tooth pair provides a single independent beating frequency. The two OFCs are ultimately combined and divided again into two paths, one of them is sent to a photodiode whose output signal is used as a reference of the mutiheterodyne signal and the other provides the light source to the PAS detection module. Figure 3 depicts the components of the dual EO comb generator and its caption specifies the commercial reference of all the components in the set-up.

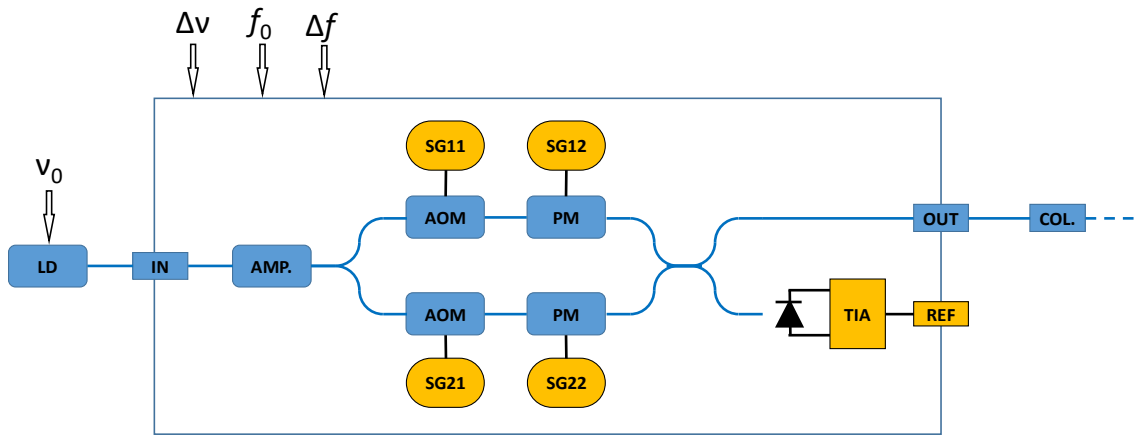

Figure 3. Dual EO comb generator set-up: (SG) RF signal generator, (IN) Input port for the laser diode (LD) with polarization control, (AMP) Optical fiber amplifier, (AOM) Acousto-optic modulator (G&H SFO2732), (PM) Ultralow Vπ phase modulators (EOSPACE PM-5K4-10/ PM-5SES-20), (TIA) Transimpedance amplifier, (REF) Comb reference output, (OUT) Optical dual comb output, (COL) Fiber collimator. All components are fiber-coupled and polarization matching is achieved by polarization maintaining components along the setup. The LD Qphotonics QDFBLD-1530-20 was used for ammonia detection.

### 3.2 Setting the dual EO comb generator for the detection module

The gas cell consists of an aluminum cylinder resonator (a pipe), with a 5-mm diameter and a length of 88 mm, terminated with two cylindrical buffer volumes as acoustic filters. Its theoretical resonance frequency cell is around 1.8 kHz [33]. To characterize the acoustic profile of the detection module, the output of the microphone placed at the center of the gas cell was connected to a lock-in amplifier (Zurich Instruments HF2LI) whose frequency reference is the same as the RF signal generators of the EO comb. We turned off the phase modulators of the dual EO comb generator so that a single tone heterodyne signal is obtained at the difference of the frequency of the AOMs ($f_0$) an we set the laser current to emit within an ammonia absorption line (around 1531.7 nm). Figure 4.a shows the amplitude of the photoacoustic signal when the parameter $f_0$ is tuned from 1700 Hz to 1950 Hz. It can be seen that the output signal depends on the acoustic response of the gas cell, showing its resonance frequency at 1820 Hz and a Q factor equal to 10. Based on this result, it is concluded that for our detection module, the parameter $f_0$ has to be set to $f_0$ = 1820 Hz and the parameter $\Delta f$ has to be set so that the bandwidth of the acoustic comb is narrower than 180 Hz. The optimum value of $\Delta f$ will depend on the desired optical resolution (number of comb teeth within the optical spectral bandwidth of interest).

Keeping the PMs turned off, we set $f_0$ = 1820 Hz (at the resonance frequency of the gas cell) and the laser current was swept from 80 mA to 105 mA, keeping a constant temperature of 26ºC. As the output power of the LD depends on the current, also does the amplitude of the heterodyne beat note used to excite the sample, so we used the reference output of the dual EO comb generator (see figure 3) to normalize the signal from the microphone. This reference signal is obtained from the second channel of the lock-in amplifier. It can be seen in Figure 4.b that the absorption profile of the ammonia is reproduced by the amplitude of the photoacoustic signal as expected [34]. This result, fitted to the reference absorption profile of the ammonia obtained from the HITRAN database allowed us to characterize the emission wavelength of the LD with its injection current. It is important to highlight that the tuning range of the LD we used in this experiment is much lower than operation wavelength range allowed by our EO dual-comb generator as it is discussed later.

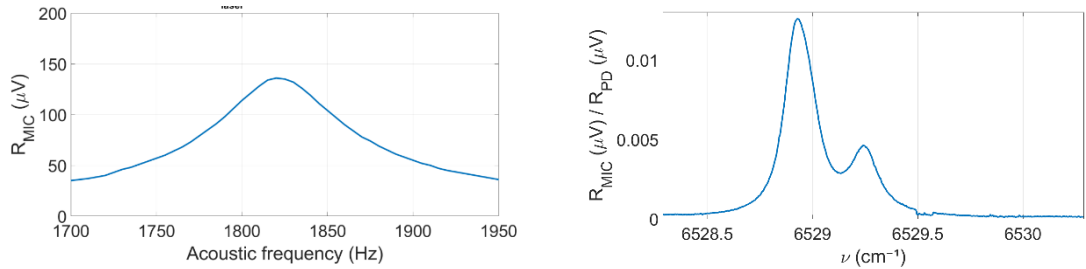

Figure 4. (a) PAS signal obtained when the difference of AOM frequencies ($f_0$) is swept, PM are turned off and laser wavelength is tuned into an absorption line. (b) PAS signal obtained when the difference of AOM frequencies ($f_0$) is 1820 Hz, PM are turned off and LD wavelength is swept as its injection current is swept.

As a reference of the performance of the set-up we compute the normalized noise equivalent absorption (NNEA): The measured optical power at the output of the fiber collimator was 0.35 mW; the measured average value of the microphone signal when the LD wavelength was tuned to the center of the absorption peak was 615 µV; the standard deviation of the microphone signal when the LD wavelength was tuned out of the absorption line was 1.5 µV, then the signal-to-noise ratio (SNR) was 396; the lock-in amplifier settings provided an equivalent bandwidth of 0.67 Hz. Given the peak absorption coefficient from the HITRAN database equal to $1.1\cdot10^{-3}$ cm$^{-1}$, the resulting NNEA was $1.2\cdot10^{-9}$ W·cm$^{-1}$·Hz$^{-1/2}$, which is a typical value for PAS.

## 4.- Results

For broadband PAS we used our dual EO comb generator. Figure 5 shows the dual comb spectra measured with an optical spectrum analyzer (OSA, Yokogawa AQ6370B). The figure shows the spectrum of two identical shape dual combs that only differ on the central wave number. These two dual comb spectra were obtained with identical configuration settings of the dual EO comb generator using a different input wavelength obtained at two different values of injection current of the LD. In this case the modulation frequency of SG11 was 5 GHz and the modulation frequency of SG21 was 5 GHz+5 Hz, both generators set with -16 dBm of RF power. These RF frequencies generate combs with a comb spacing of $\Delta v$ = 5 GHz (equivalent to 0.167 $cm^{-1}$). In this figure we can observe that the dual comb generator generates identical combs for given PMs parameters with a central frequency depending on the LD seed wavelength. Given the modulation depth (RF power), we obtain seven dual comb teeth within a -20 dB bandwidth, giving a spectral coverage of 30 GHz (1 $cm^{-1}$) with a resolution of 5 GHz (0.167 $cm^{-1}$).

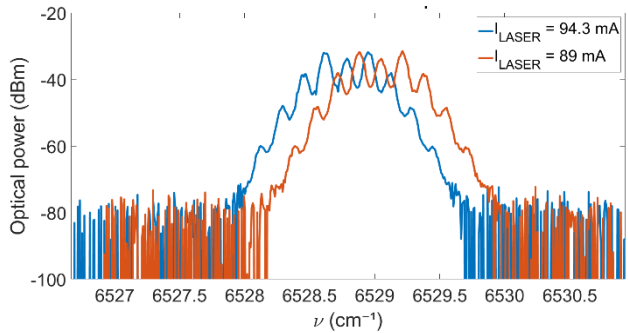

Figure 5. Dual comb generated with the following configuration parameters: $f_0$ = 1820 Hz, $\Delta v$ = 5 GHz, $\Delta f$ = 5 Hz: The frequency of the RF generators was 5 GHz (SG12), 5 GHz + 5 Hz (SG22), 40 MHz (SG11) and 40 MHz + 1820 Hz (SG21). The center of the spectrum $v_0$ depends on the LD emission wavelength that depends on its injection current: 94.3 mA (blue), and 89 mA (red).

As stated before, the difference of the modulation frequency of SG11 and SG22 is $\Delta f$ = 5 Hz. With these setting the spectral compression ratio $\Delta v/\Delta f$ equals $10^9$, so that the bandwidth of the resulting acoustic comb is 30 Hz, small enough to fit with-in the bandwidth of our gas cell having set the difference of the frequencies of the AOMs (SG21 and SG22) to the resonance frequency of the gas cell ($f_0$ = 1820 Hz). To recover the absorption profile of the gas sample when we use the dual EO comb as excitation source we followed the next steps. First, we measured the amplitude of the microphone signal with the lock-in amplifier at the frequencies $f_0$ + n $\Delta f$ (in this case 1805 Hz, 1810 Hz, 1815 Hz, 1820 Hz, 1825 Hz, 1830 Hz and 1835 Hz), then the amplitude of each acoustic mode is corrected taking into account the gas cell resonance profile obtained at calibration and represented in figure 4.a. Second, the amplitude of each mode is normalized with the corresponding reference comb obtained from the photodiode signal. Finally, each acoustic frequency is mapped to the optical domain given the EO comb generator parameters. Figures 6 and 7 represent the results for the dual combs in figure 5 along with the reference absorption profile obtained by LD current swept (figure 4.b). It can be seen that the generated acoustic comb reproduced the absorption profile of the gas sample with a resolution given by the dual-comb source.

Given the versatility of the EO comb we are able to increase the optical resolution of the measurements by reducing the modulation frequency of the PMs. It is also possible to generate a higher number of modes within the same spectral bandwidth increasing the modulation depth of the PMs. As example, figure 8.a shows the reference comb when the RF power applied to the PMs is increased from -16 dBm (RF power applied for the results represented in figures 5 to 7) to -10 dBm. It can be seen that the amplitude of the comb is reduced as the available total optical power remains constant. As a tradeoff between spectral resolution and SNR, which mainly depends on the optical power at each frequency, we have considered that 10 optical frequencies within the absorption feature offers enough spectral resolution to reconstruct its shape. Figure 8 shows the DCPAS results when the dual EO comb generator is configured to obtain this optimal spectral resolution, the configuration parameters are detailed in the figure caption.

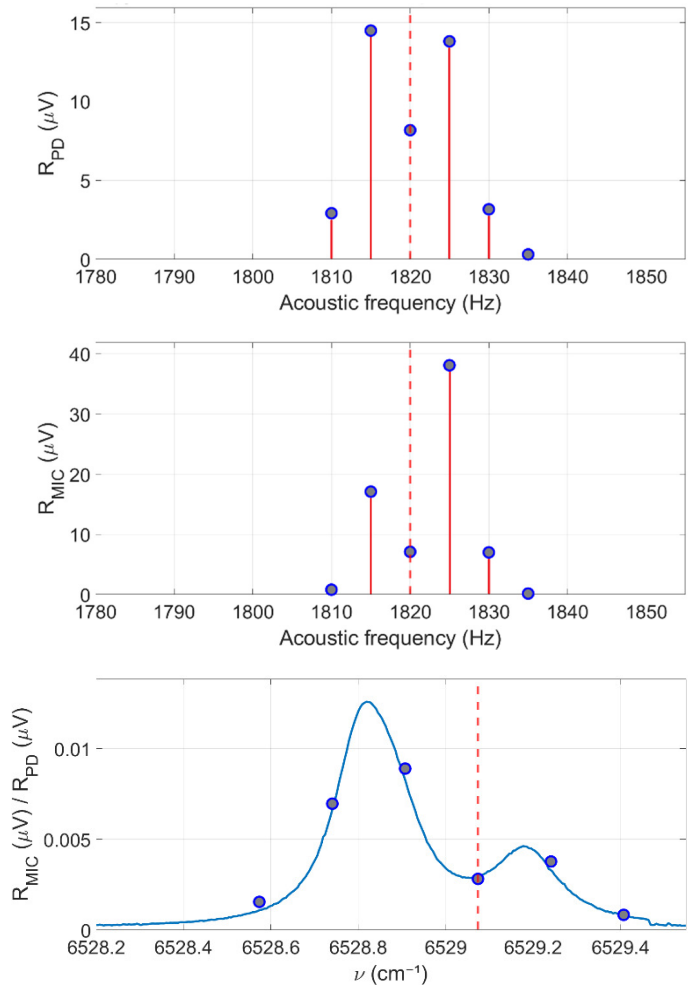

Figure 6. DCPAS results with the following configuration parameters: $f_0$ = 1820 Hz, $\Delta\nu$ = 5 GHz and $\Delta f$ = 5 Hz. (a) The dots are the amplitudes of the reference heterodyne signals measured at the reference photodiode of the EO comb generator. (b) The dots represent the amplitudes of the tones of the photoacoustic signal detected by the microphone. (c) The dots represent the normalized photoacoustic signal after optical - acoustic mapping: In this case $\nu_0$ = 6529.07 cm$^{-1}$ ($I_{LASER}$ = 89 mA) and $\Delta\nu$ = 0.17 cm$^{-1}$ (5 GHz). The measured absorption profile of ammonia is represented as reference.

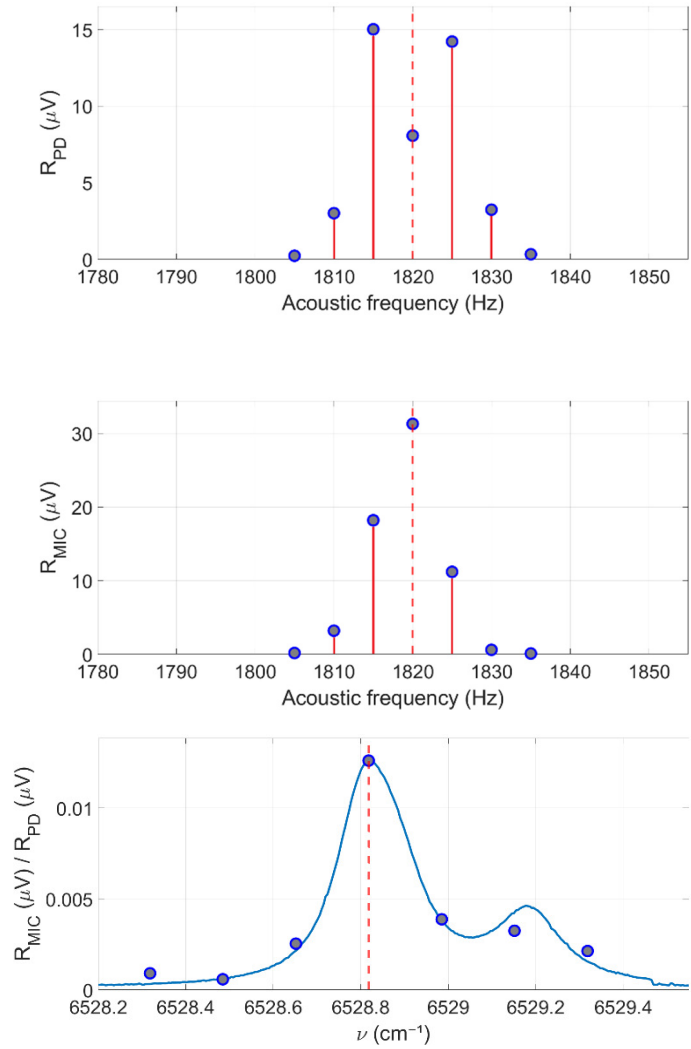

Figure 7. DCPAS results with the following configuration parameters: $f_0$ = 1820 Hz, $\Delta\nu$ = 5 GHz and $\Delta f$ = 5 Hz. (a) The dots are the amplitudes of the reference heterodyne signals measured at the reference photodiode of the EO comb generator. (b) The dots represent the amplitudes of the tones of the photoacoustic signal detected by the microphone. (c) The dots represent the normalized photoacoustic signal after optical - acoustic mapping: In this case $\nu_0$ = 6528.82 cm$^{-1}$ ($I_{LASER}$ = 94.3 mA) and $\Delta\nu$ = 0.17 cm$^{-1}$ (5 GHz). The measured absorption profile of ammonia is represented as reference.

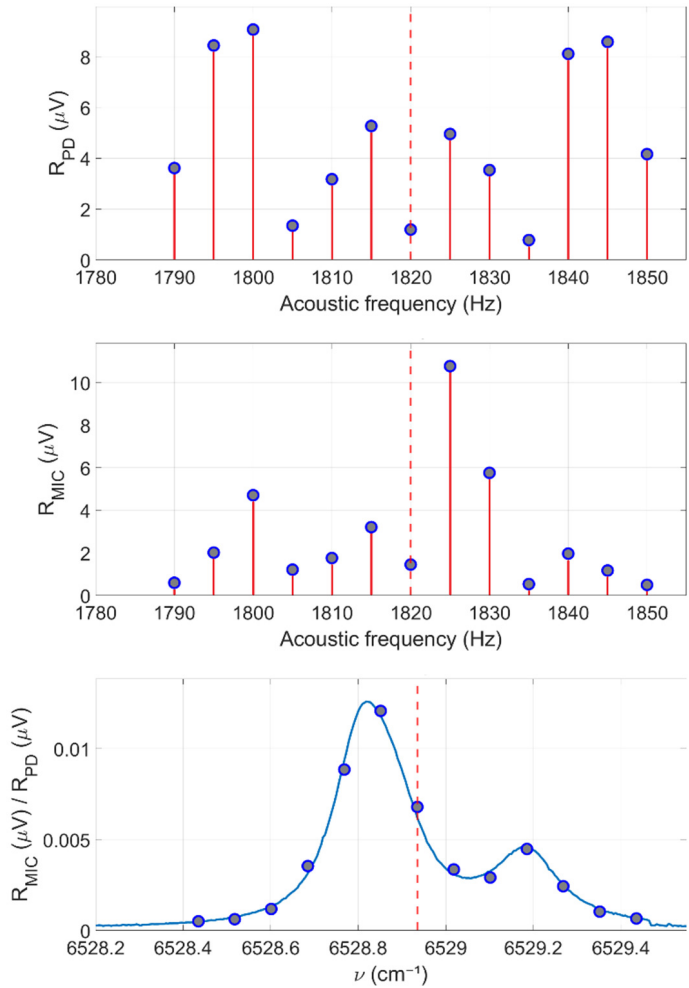

Figure 8. DCPAS results with the following configuration parameters: $f_0$ = 1820 Hz, $\Delta v$ = 2.5 GHz and $\Delta f$ = 5 Hz. (a) Reference comb (b) Photoacoustic tones (c) Normalized photoacoustic signal after optical - acoustic mapping: $v_0$ = 6528.86 cm$^{-1}$ and $\Delta v$ = 0.08 cm$^{-1}$ (2.5 GHz). The measured absorption profile of ammonia is represented as reference.

## 5. Discussion and conclusion:

In this paper we have demonstrated a DCPAS set-up for ammonia detection in the near-infrared using an electro optic dual comb generator. The proposed method permits broadband PAS virtually with any standard or custom detection module and the simultaneous interrogation of the whole spectral range with an easily configurable (very high) optical resolution.

In our system, each comb is generated with a single PM whose input optical frequency is slightly shifted with an AOM. This simple architecture and particular generation method provide us with the ability to spectrally interrogate an absorption signature of the target gas not only, as introduced above, with variable resolution but, most importantly, to perform the optical to acoustic mapping so that all the acoustic modes are generated with-in the bandwidth of the photoacoustic detection module.

One of the main features of the dual comb system employed, is that the two combs generated by the EO modulators typically exhibit relatively narrow spectral bandwidth compared to other complex comb platform. This is an advantage for PAS since the available optical power per teeth can be maximized, thus optimizing the signal-to-noise ratio without the need of optical filtering. Compared to PAS system that use a single optical frequency to excite the sample, when the absorption line is sampled at several frequencies, even if the center frequency of the laser may shift, the shape of the line can be determined from the spectrum of the acoustic signal. This relaxes the need of laser frequency stabilization and the need to control the pressure and temperature of the sample. Also interfering absorptions can be identified.

In this work we have demonstrate the DCPAS technique in the near infrared, were EO components are readily available, their technological maturity is very high and all fiber optic components can be used. Recent advances on EO comb generation architectures in the near infrared, also relevant for PAS, focus on increasing compactness, spectral bandwidth and mutual coherence [11][35]. It is also worth mentioning that the greatest potential of this technique is its direct application in any other region of the spectrum thanks to the indirect measurement of the absorption through the photoacoustic response. In fact, the current implementation of the dual EO comb generator can operate over a wide range of wavelengths and other EO architectures can be used to operate from the visible to the mid infrared [36]. Besides this, frequency conversion techniques can also be employed to directly and easily shift the operation range of EO comb generators to the, extremely interesting for gas detection and identification, mid infrared range [37].

The system whose operation has been analyzed in the previous paragraphs has been experimentally validated for the reconstruction of the absorption profile of ammonia around 6529 $cm^{-1}$ wavenumbers within a spectral bandwidth of 1 $cm^{-1}$ and variable resolution, finding 2.5 GHz (0.08 $cm^{-1}$) a good compromise between spectral resolution and power per comb line. We have also validated the dual EO comb generation technique to easily match the frequencies of the multiherodyne tones within the band of resonance of the gas cell, in this case a tube resonator with resonance frequency 1820 Hz and Q factor equal to 10. The results obtained corroborate not only the technical feasibility of the method, but also the great potential it may have as future developments that incorporate more sophisticated photoacoustic detection modules.

## Acknowledgements

This work is supported by the State Research Agency of Spain under grant PID2020-116439GB-I00 and by Madrid Government under personnel grant PEJ-2020-AI/TIC-19407.